\documentclass[twocolumn]{aastex63}

\usepackage{rotating}
\usepackage{acronym}
\usepackage{xspace}
\usepackage{multirow}
\usepackage{mathtools}
\usepackage{amsmath}
\usepackage{hyperref}
\usepackage{units}

\usepackage{xcolor}

\graphicspath{{./}{figures/}}

\definecolor{chmagenta}{rgb}{0.54, 0.17, 0.88}

\def\q{\ensuremath{q}\xspace}
\def\z{\ensuremath{z}\xspace}

\def\Msun{\ensuremath{\mathit{M_\odot}}\xspace}
\def\Rsun{\ensuremath{\mathit{R_\odot}}\xspace}

\def\Pcc{\ensuremath{\mathit{P_\mathrm{cc}}}\xspace}
\def\tmin{\ensuremath{\mathit{t_\mathrm{min}}}\xspace}
\def\tmax{\ensuremath{\mathit{t_\mathrm{max}}}\xspace}

\def\tdel{\ensuremath{\mathit{t_\mathrm{d}}}\xspace}
\def\tinsp{\ensuremath{\mathit{t_\mathrm{insp}}}\xspace}
\def\tstar{\ensuremath{\mathit{t_\star}}\xspace}
\def\tmin{\ensuremath{\mathit{t_\mathrm{min}}}\xspace}
\def\tmax{\ensuremath{\mathit{t_\mathrm{max}}}\xspace}

\def\NPhotoDatapoints{\ensuremath{542}\xspace}
\def\NSpectra{\ensuremath{42}\xspace}
\def\NOffsets{\ensuremath{83}\xspace}

\def\NHostTotal{\ensuremath{84}\xspace}
\def\NHostThreePhotometricBands{\ensuremath{69}\xspace}
\def\NHostSpecz{\ensuremath{71\%}\xspace}

\def\NHost{\ensuremath{68}\xspace}
\def\NHostSpec{\ensuremath{48}\xspace}
\def\NHostGold{\ensuremath{39}\xspace}
\def\NHostSpecGold{\ensuremath{29}\xspace}
\def\NhostStarFormingPercentage{\ensuremath{\approx 82\%}\xspace}
\def\NhostTransitioningPercentage{\ensuremath{\approx 6\%}\xspace}
\def\NhostQuiescentPercentage{\ensuremath{\approx 12\%}\xspace}

\def\AllSamplesAlpha{\ensuremath{-1.83^{+0.35}_{-0.39}}\xspace}
\def\AllSamplesAlphaNinetyNinethPercentile{\ensuremath{-1.31}\xspace}
\def\AllSamplesTmin{\ensuremath{184^{+67}_{-79}~\mathrm{Myr}}\xspace}
\def\AllSamplesTminFirstPercentile{\ensuremath{72~\mathrm{Myr}}\xspace}
\def\AllSamplesTmaxFirstPercentile{\ensuremath{7.95~\mathrm{Gyr}}\xspace}

\def\SpeczSamplesAlpha{\ensuremath{-1.65^{+0.39}_{-0.38}}\xspace}

\def\SpeczSamplesTminFirstPercentile{\ensuremath{88~\mathrm{Myr}}\xspace}
\def\DifferenceAllSamplesSpeczTminApproximate{\ensuremath{20~\mathrm{Myr}}\xspace}

\def\TmaxExtendedPeakApproximate{\ensuremath{\sim 13~\mathrm{Gyr}}\xspace}

\def\MergerRatePeakRedshiftApproximate{\ensuremath{1.6\mbox{--}1.7}\xspace}
\def\LocalMergerRateApproximate{\ensuremath{135~\mathrm{Gpc}^{-3}~\mathrm{yr}^{-1}}\xspace}

\def\MinimumSeparationTminFirstPercentile{\ensuremath{1.1~R_\odot}\xspace}
\def\SeventeenZeroEightSeventeenExclusionChange{\ensuremath{5\%}\xspace}

\acrodef{GW}{gravitational wave}
\acrodef{BH}{black hole}
\acrodef{BBH}{binary black hole}
\acrodef{NS}{neutron star}
\acrodef{BNS}{binary neutron star}
\acrodef{EM}{electromagnetic}

\acrodef{LIGO}{Laser Interferometer Gravitational-wave Observatory}
\acrodef{LVC}{LIGO Scientific and Virgo Collaboration}
\acrodef{O1}{first observing run}
\acrodef{O2}{second observing run}
\acrodef{O3}{third observing run}
\acrodef{O3a}{first half of the third observing run}

\acrodef{DTD}{delay time distribution}

\acrodef{GRB}{gamma-ray burst}

\acrodef{SFH}{star formation history}
\acrodef{SFR}{star formation rate}

\newcommand{\KICP}{\affiliation{Kavli Institute for Cosmological Physics, The University of Chicago, 5640 South Ellis Avenue, Chicago, IL 60637, USA}}
\newcommand{\EFI}{\affiliation{Enrico Fermi Institute, The University of Chicago, 933 East 56th Street, Chicago, IL 60637, USA}}
\newcommand{\UChicago}{\affiliation{Department of Physics, Department of Astronomy \& Astrophysics, The University of Chicago, 5640 South Ellis Avenue, Chicago, IL 60637, USA}}
\newcommand{\IISER}{\affiliation{Department of Physics, Indian Institute of Science Education and Research, Homi Bhaba Road, Pune, Maharashtra, 411045, India}}
\newcommand{\NU}{\affiliation{Center for Interdisciplinary Exploration and Research in Astrophysics (CIERA) and Department of Physics and Astronomy, Northwestern University, Evanston, IL 60208, USA}}

\shorttitle{Delay Time Distribution of Short GRBs}
\shortauthors{Zevin et al. 2022}

\begin{document}

\title{Observational Inference on the Delay Time Distribution of Short Gamma-ray Bursts}

\author[0000-0002-0147-0835]{Michael Zevin}\email{michaelzevin@uchicago.edu}\thanks{NASA Hubble Fellow}
\KICP \EFI
\author[0000-0002-2028-9329]{Anya E. Nugent}
\NU
\author[0000-0002-0298-4432]{Susmita Adhikari}
\UChicago \IISER 
\author[0000-0002-7374-935X]{Wen-fai Fong}
\NU
\author[0000-0002-0175-5064]{Daniel E. Holz}
\KICP\EFI\UChicago
\author[0000-0002-6625-6450]{Luke Zoltan Kelley}
\NU

\begin{abstract}
The delay time distribution of neutron star mergers provides critical insights into binary evolution processes and the merger rate evolution of compact object binaries. 
However, current observational constraints on this delay time distribution rely on the small sample of Galactic double neutron stars (with uncertain selection effects), a single multimessenger gravitational wave event, and indirect evidence of neutron star mergers based on $r$-process enrichment. 
We use a sample of \NHost host galaxies of short gamma-ray bursts to place novel constraints on the delay time distribution and leverage this result to infer the merger rate evolution of compact object binaries containing neutron stars. 
We recover a power-law slope of $\alpha = \AllSamplesAlpha$ (median and 90\% credible interval) with $\alpha < \AllSamplesAlphaNinetyNinethPercentile$ at 99\% credibility, a minimum delay time of $\tmin = \AllSamplesTmin$ with $\tmin > \AllSamplesTminFirstPercentile$ at 99\% credibility, and a maximum delay time constrained to $\tmax > \AllSamplesTmaxFirstPercentile$ at 99\% credibility. 
We find these constraints to be broadly consistent with theoretical expectations, although our recovered power-law slope is substantially steeper than the conventional value of $\alpha = -1$, and our minimum delay time is larger than the typically assumed value of $10~\mathrm{Myr}$. 
Pairing this cosmological probe of the fate of compact object binary systems with the Galactic population of double neutron stars will be crucial for understanding the unique selection effects governing both of these populations. 
In addition to probing a significantly larger redshift regime of neutron star mergers than possible with current gravitational wave detectors, complementing our results with future multimessenger gravitational wave events will also help determine if short gamma-ray bursts ubiquitously result from compact object binary mergers. 
\end{abstract}


\section{Introduction}\label{sec:intro}

The transient event GW170817~\citep{GW170817}, which was observed in both \acp{GW} and \ac{EM} waves, firmly established the connection between \ac{BNS} mergers and short \acp{GRB}. 
The host galaxy of this event, NGC4993, was subsequently identified through broadband \ac{EM} emission of the ensuing kilonova~\citep{Allam2017,Arcavi2017,Coulter2017a,Lipunov2017,Tanvir2017a,Yang2017} and eventually, the nonthermal afterglow of the \ac{GRB}~\citep[see][and references therein]{Margutti2021}. 
Under the paradigm that all short \acp{GRB} originate from compact object binary mergers, such host associations can unveil unprecedented information regarding the stellar populations in galaxies that host compact object binary mergers, thereby allowing for novel constraints on the progenitors of these events. 

One key aspect of compact object binary mergers that can be illuminated using such host associations is the \ac{DTD}. 
The \ac{DTD} yields important information regarding the birth properties of compact object binaries, such as their typical orbital separations at birth and inspiral times. 
This, in turn, provides critical constraints on evolutionary processes of the progenitor binary stellar system and can inform modeling of massive-star binaries, which form \ac{GW} sources accessible by current and future observatories. 

GW170817 is the only \ac{GW} event to date where the host galaxy has been confidently established, and a number of fortuitous aspects of the system that produced this signal were paramount in the discovery of its \ac{EM} counterpart: for example, it was well within the horizon of the \ac{GW} detectors~\citep{LVC_ObservingScenarios}, the \ac{GRB} jet was aligned $\approx20^\circ$ from our line of sight and eventually came into view \citep{Margutti2021}, and it was in close proximity to a massive galaxy that was prioritized by targeted searches \citep{Coulter2017}, although wide-field follow-up searches would have found it regardless~\citep{Soares-Santos2017}. 
While the \ac{BNS} detection rate in the forthcoming observing run of the LIGO--Virgo--Kagra interferometer network will be $\mathcal{O}(1)$/month~\citep{LVC_ObservingScenarios}, the rate of \ac{BNS} mergers with an \ac{EM} counterpart detection precise enough for confident host association is more uncertain given the unique observational challenges~\citep[e.g.,][]{Coughlin2018,Dichiara2020,Colombo2022,Perna2022}. 
In the next observing run, the \ac{BNS} range for the LIGO detectors is expected to be $160\mbox{--}190~\mathrm{Mpc}$~\citep{LVC_ObservingScenarios}, and \ac{GW} mergers are more likely to be detected close to the detector horizon where the sensitive volume is the largest. 
Three-dimensional \ac{GW} sky localization volumes will be larger at these distance, and even with improved search capabilities from state-of-the-art wide-field telescopes, it is unclear how rapidly multimessenger events with definitive host associations will be accumulated in the coming years. 

On the other hand, the number of short \acp{GRB} with confident host galaxy associations has increased substantially over the past two decades. 
Over the next decade, the cumulative number of these host associations may still exceed the number of host associations found via multimessenger observations. 
\ac{GRB} monitoring missions such as Swift and Fermi have observed hundreds of short \acp{GRB}, with subsequent searches in the X-ray, optical, and radio bands for \ac{GRB} afterglows leading to the identification of probable host galaxies for dozens of these events over a wide range of redshifts \citep{Berger2014,Fong2022,Nugent2022,OConnor2022}. 
Photometric and spectroscopic follow-up observations of the afterglow and the host galaxy can then unveil crucial aspects of the \ac{GRB}--host connection that encode information about the progenitors of \ac{BNS} systems themselves, such as the stellar mass and \ac{SFR} of the galaxy, its \ac{SFH}, and the galactocentric offset of the \ac{GRB}~\citep{Fong2013,Fong2013a}. 

Here, we consider direct implications of \ac{GRB}--galaxy connections on compact object binary formation and the evolution of their progenitor stars, leveraging an updated sample of \ac{GRB} host galaxies and their properties presented in \cite{Fong2022} and \cite{Nugent2022} to place novel constraints on the \ac{DTD} of short \acp{GRB}. 
This expansive catalog of short \ac{GRB} hosts is likely to outnumber the hosts of multimessenger \ac{GW} events for years to come, and depending on the highly uncertain rate of discovery of \ac{EM} counterparts to \ac{GW} events, may not be met with multimessenger events until the onset of third-generation \ac{GW} detectors. 
Besides probing a much larger cosmological volume than a multimessenger sample, future comparisons of \ac{DTD} constraints between short \acp{GRB} and \acp{GW} with \ac{EM} counterparts could help determine whether the \ac{BNS} merger paradigm for short \acp{GRB} is universal. 
Furthermore, \ac{DTD} constraints from such an extragalactic population are complementary to those from compact object binaries observed in the Milky Way and with \acp{GW}, allowing for comparisons of these distinct probes to help unravel selection effects that impinge upon the detection of each population individually. 

The format of this Letter is as follows: 
In Section~\ref{sec:grb_hosts}, we summarize the catalog of \ac{GRB} host galaxies used in our analysis and assumptions regarding their \acp{SFH}. 
Our main results are in Section~\ref{sec:delay_time_distributions}, where we describe our inference methodology and present our constraints on the \ac{DTD} of short \acp{GRB}. 
We show implications of our constraints on the predicted compact object binary merger rate in Section~\ref{sec:merger_rate_evolution}.  
In Section~\ref{sec:discussion} we discuss caveats of our analysis and highlight other aspects of binary stellar evolution and compact object binary formation that will be addressed in future work. 
We summarize our results in Section~\ref{sec:conclusions}.
Throughout this work we employ a standard \texttt{WMAP} cosmology of $H_{0}$ = 69.6~km~s$^{-1}$~Mpc$^{-1}$, $\Omega_{M}$ = 0.286, $\Omega_{\Lambda}$ = 0.714 \citep{Hinshaw2003,Bennett2014}, consistent with \cite{Nugent2022}. 
All code and data used in this analysis are available on Zenodo.\footnote{\url{https://doi.org/10.5281/zenodo.7015221}}

\section{Catalog of Short GRB Hosts}\label{sec:grb_hosts}

Recently, \cite{Fong2022} presented an updated catalog of \NHostTotal short \ac{GRB} hosts with broadband photometry and well-studied explosion environments. 
Combined with the literature, the catalog comprises \NPhotoDatapoints photometric data points, \NSpectra spectra, and \NOffsets offset measurements, and is comprehensive for Swift short \acp{GRB} and neutron star \ac{GW} merger events discovered in 2005--2021.\footnote{The sample in \cite{Fong2022} also includes long-duration GRB060614 and GRB211211A, which are thought to be the result of neutron star mergers due to the lack of a supernova to deep optical limits and the observation of a kilonova, respectively~\citep{Gehrels2006,Rastinejad2022}.}
This catalog, as well as other recent short \ac{GRB} catalogs~\citep[e.g.,][]{OConnor2022}, have more than doubled the number of short \acp{GRB} with confident host associations and also include a number of systems that are found to be ``hostless'' with no coincident host to deep optical limits, depending on the threshold defining host galaxy association~\citep{Berger2010}. 

Of the \NHostTotal short \ac{GRB} hosts with broadband photometry presented in \cite{Fong2022}, \NHostThreePhotometricBands were determined to have sufficient observational data (detected in $\geq 3$ photometric bands) to model the galactic spectral energy distribution and infer aspects of the host; the detailed properties of this population of short \ac{GRB} host galaxies are analyzed in \cite{Nugent2022}. 
Approximately \NHostSpecz of these hosts have spectroscopic redshifts. 
For the main analysis in this work, we include in our sample NGC4993, the host galaxy of multimessenger event GW170817/GRB170817, and do not include the long-duration GRB060614 and GRB211211A despite that they may be the result of compact object binary mergers; we perform additional fits to the \ac{DTD} with the inclusion of the host galaxies of these two long-duration \ac{GRB} events and comment on the impact in Section~\ref{sec:discussion}. 
This totals in \NHost host galaxies used in our main analysis, with \NhostStarFormingPercentage of the hosts being star-forming, \NhostTransitioningPercentage transitioning, and \NhostQuiescentPercentage quiescent. 
The properties of the host galaxies in our sample are modeled consistently as described below.

\subsection{Host Galaxy Properties}\label{subsec:star_formation_histories}

Properties of the host galaxies, such as host redshift, stellar mass, stellar population ages, metallicity, and \ac{SFR}, are inferred using the \texttt{Prospector} package~\citep{Leja2017,Johnson2021}, as described in \cite{Nugent2022}. 
\texttt{Prospector} uses the \texttt{dynesty} nested sampler~\citep{Speagle2020} to constrain aspects of galaxies based on available photometric and spectroscopic data. 
Redshifts are fixed in this inference for galaxies that have spectroscopically determined redshifts, as the redshift measurement uncertainties for these hosts are negligible. 

The \ac{SFH} of each host, $\psi(t)$, is modeled assuming a delayed-$\tau$ functional form,\footnote{
A number of other parametric functional forms~\citep{Carnall2019} or nonparametric approaches~\citep{Leja2019} can be used when reconstructing the \ac{SFH}, sometimes yielding discrepant results. However, parametric fits are used in this work to better establish uniformity in our dataset, which has an inconsistent amount and quality of data across host observations. 
Parametric SFHs are also more commonly used in host galaxy literature, making direct comparisons more seamless.
We comment on this systematic uncertainty in Section~\ref{sec:discussion}. } 
\begin{equation}\label{eq:delayed_tau}
    \psi(t) = M_\mathrm{F} \times 
    \frac{t e^{-t/\tau}}{\int_0^{t_\mathrm{SF} } t^\prime e^{-t^\prime/\tau} dt^\prime},
\end{equation}
where $M_\mathrm{F}$ is the total stellar mass formed in the galaxy, $t_\mathrm{SF}$ is the time at which star formation commenced relative to the observation time, and $\tau$ is the e-folding factor. 
Posterior samples for the parameters $M_\mathrm{F}$, $t_\mathrm{SF}$, and $\tau$, as well as the redshift $z$ for hosts that only have photometric redshifts, fully construct the \ac{SFH} for each \texttt{Prospector} posterior sample. 
We convert Eq.~\ref{eq:delayed_tau} into a redshift-dependent \ac{SFH}, $\psi(z)$, in our \ac{DTD} parameter hyperlikelihood described in Section~\ref{subsec:inference}.

\subsection{Criteria for Host Associations}\label{subsec:criteria_host_associations}

The probability of chance coincidence (\Pcc) metric is used for determining the confidence of a particular host association~\citep{Bloom2002}. 
This metric decreases (i.e., results in a higher likelihood that the \ac{GRB} is correctly associated with a particular host) with decreasing optical magnitude of the potential host and decreasing angular offset of the \ac{GRB}, which leads to a higher confidence in host association for \acp{GRB} in close proximity to higher-mass galaxies. 
We use the \Pcc criteria of \cite{Fong2022} to broadly categorize \acp{GRB} in the catalog based on host association confidence: a ``Gold Sample'' with $\Pcc \leq 0.02$, a ``Silver Sample'' with $0.02 < \Pcc \leq 0.10$, and a ``Bronze Sample'' with $0.10 < \Pcc \leq 0.20$.

In this work, our full sample considers the \NHost hosts observed in $\geq 3$ photometric bands with $\Pcc \leq 0.20$. 
We also present results using three subsets of this sample based on their redshift uncertainty and/or \Pcc: a subset of hosts with spectroscopic redshifts (and therefore negligible redshift uncertainty; \NHostSpec hosts), a subset of hosts with $P_{cc} \leq 0.02$ (Gold Sample; \NHostGold hosts), and a subset of hosts in the Gold Sample that also have spectroscopic redshifts (\NHostSpecGold hosts). 
Note that in our main analysis we include GRB170817, which falls under the Gold Sample and has a spectroscopic redshift, and do not include long-duration GRB060614 and GRB211211A; hence the slight differences in the number of hosts in our subsamples compared to \citealt{Nugent2022}; see their Figure 1. 
We comment on completeness and potential biases from excluding certain \ac{GRB} hosts from our sample in Section~\ref{sec:discussion}.

\section{Delay Time Distributions}\label{sec:delay_time_distributions}

The delay time of compact object binaries is defined as the time between the formation of the progenitor stars and the merger of the two compact objects, such that $\tdel = \tstar + \tinsp$ where $\tstar$ is the time between stellar birth and compact object binary formation, and $\tinsp$ is the compact object binary inspiral time.
The \ac{DTD} for isolated compact binaries is typically parameterized as a power-law distribution\footnote{Another commonly used functional form for the \ac{DTD} of short \acp{GRB} is a log-normal distribution~\citep{Nakar2006,Berger2007,Wanderman2015}, though this functional form is in tension with the growing number of high-redshift short \acp{GRB} and is not consistent with the expected \ac{DTD} from compact object binary mergers based on population modeling predictions and the power-law orbital period distribution of their binary massive-star progenitors.} with a minimum delay time $\tmin$, which encodes the minimum stellar evolutionary timescales required to form a compact object binary ($\mathcal{O}(10~\mathrm{Myr})$ for \ac{BNS} progenitors) and the minimum orbital separation of a compact object binary at formation, and can include a (somewhat arbitrary) maximum delay time $\tmax$ that can be much larger than the Hubble time,
\begin{equation}
p(\tdel|\alpha,\tmin,\tmax) =
\begin{cases}
\mathcal{N} \tdel^{\alpha}, & \tmin \leq \tdel \leq \tmax\\
0, & \text{otherwise}\\
\end{cases}
\end{equation}
where $\alpha$ is the spectral index of the power-law distribution and
\begin{equation}
\mathcal{N} =
\begin{cases}
\left[ \mathrm{log}_{10} \left( \frac{\tmax}{\tmin} \right) \right]^{-1}, & \alpha = -1\\
(1 + \alpha)(\tmax^{1+\alpha} - \tmin^{1+\alpha})^{-1}, & \text{otherwise}\\
\end{cases}
\end{equation}
is the normalization constant for the probability distribution function. 

The slope of the \ac{DTD} for compact object binary mergers is typically assumed to be $\alpha \approx -1$~\citep[e.g.,][]{Piran1992}. 
This assumption stems from the fact that in general $\tinsp \gg \tstar$ and from coupling the equations that govern the \ac{GW} inspiral with an assumption that the orbital separation distribution of compact object binaries follows the distribution of massive O/B stars, which is approximately flat in log with $dN/da \propto a^{-1}$ where $a$ is the semimajor axis (for example, the best-fit orbital separation distribution from a survey of massive-star binary initial properties in \citealt{Sana2012} finds $dN/da \propto a^{-0.83}$). 
Because the inspiral time scales as $t_\mathrm{insp} \propto a^4$, it follows that $da/dt_\mathrm{insp} \propto t_\mathrm{insp}^{-3/4}$ and, assuming $dN/da \propto a^{-1}$, $dN/da \propto t_\mathrm{insp}^{-1/4}$. 
The distribution of inspiral times is thus $dN/dt_\mathrm{insp} \propto t_\mathrm{insp}^{-1}$. 
However, hardening phases during the coevolution of massive-star binaries, such as common envelopes or stable mass transfer phases from a more massive donor star to a less massive compact object accretor, can lead to steeper orbital separation distributions and therefore steeper slopes in the \ac{DTD}~\citep[e.g.,][]{Belczynski2018}. 
The \ac{DTD} inferred from short \acp{GRB} is thus useful in constraining a variety of uncertain aspects of massive-star binary evolution. 

Multiple observational probes have been explored in constraining aspects of the short \ac{GRB}/\ac{BNS} \ac{DTD}. 
The redshift distribution of short \acp{GRB} has been able to place broad constraints on the \ac{DTD}~\citep{Nakar2006,Berger2007,Jeong2010,Hao2013,Wanderman2015,Anand2018}, with high-redshift \acp{GRB} in particular allowing for stronger constraints on the minimum delay time and power-law slope of the \ac{DTD}~\citep{Paterson2020,Nugent2022,OConnor2022}. 
The binary--host connection interpreted via host galaxy demographics and galaxy scaling relations has also shown promise in constraining \ac{BNS} kick velocities, delay times, and the properties of galaxies that host \ac{BNS} mergers and short \acp{GRB}~\citep{Zheng2007,Kelley2010,Fong2013,Behroozi2014,Adhikari2020,Santoliquido2022}. 
The Milky Way offers a limited sample of \ac{BNS} systems with well-characterized orbital properties and inspiral times, many of which are much longer than the Hubble time, that are useful for examining the \ac{BNS} \ac{DTD}~\citep{Beniamini2016,Tauris2017,Vigna-Gomez2018,Andrews2019,Beniamini2019}. 
However, uncertain selection effects inherent to this sample may affect inference of \ac{DTD} parameters~\citep[e.g.,][]{Tauris2017}. 
All these observational probes are complemented by population modeling of compact object binaries that explore variations in binary evolution physical assumptions, and therefore the predicted \ac{DTD} of compact object binaries~\citep{Belczynski2018,Chruslinska2018,Broekgaarden2022a,Santoliquido2022}; these predicted compact object populations can be compared and constrained with the population properties and merger rates observed by \ac{GW} detectors~\citep{GWTC3}. 
Though \ac{GW} observations of compact object binary mergers also hold promise in constraining the \ac{DTD} of compact object mergers, both through the use of galaxy scaling relations~\citep{Safarzadeh2019,Safarzadeh2019a} and multimessenger events whose host galaxies have been identified and characterized~\citep{Safarzadeh2019d}, it may require the accumulation of hundreds of host galaxy identifications to achieve strong constraints on the \ac{DTD}. 

\subsection{Inferring the DTD Parameters}\label{subsec:inference}

We follow \cite{Safarzadeh2019d} in constructing our likelihood function for the \ac{DTD}, which is the key component of our population inference. 
From the \texttt{Prospector} modeling of each host galaxy, we have a set of posterior samples that define the \ac{SFH} and, for hosts without spectroscopically measured redshifts, the galaxy redshift; a fixed redshift is used for the galaxies that have spectroscopic redshifts. 
We draw $N_\mathrm{samp} = 100$ \texttt{Prospector} samples from each host, where we weight our draws with the inverse of the prior to get draws from the likelihood rather than the posterior.\footnote{All relevant \texttt{Prospector} parameters used in this analysis were sampled with uniform priors, making this prior reweighting irrelevant.}
For a given host galaxy $i$ and likelihood sample $j$, the expected merger rate at the measured redshift $z_i^j$ is 
\begin{equation}\label{eq:merger_rate}
    \dot{n}_i^j = \int_{z^\prime=\infty}^{z^\prime=z_i^j} p(t^\prime_\mathrm{lb}-\tdel | \alpha, \tmin, \tmax) \lambda \psi_i^j(z^\prime)\frac{dt}{dz}(z^\prime) dz^\prime,
\end{equation}
where $t^\prime_\mathrm{lb}$ is the lookback time at redshift $z^\prime$; $\lambda$ is the \ac{BNS} formation efficiency which we assume is $10^{-5} \Msun^{-1}$ and does not evolve with redshift~\citep[see e.g.,][]{Broekgaarden2022a}; $\psi(z)$ is the \ac{SFR} as a function of redshift defined in Section~\ref{subsec:star_formation_histories}; $dt/dz = [H_0 E(z) (1+z)]^{-1}$; and $E(z) = \sqrt{\Omega_{M}(1+z)^3 + \Omega_{\Lambda}(z)}$. 

The probability of observing a single short \ac{GRB} in a host galaxy over a set interval of time $\Delta t$ follows a Poisson distribution based on the merger rate of Equation~\ref{eq:merger_rate}. 
Approximating the marginalization over likelihood samples as a discrete sum, the hyperlikelihood for observing a single short \ac{GRB} in a particular host galaxy given the set of \ac{DTD} parameters becomes 
\begin{equation}\label{eq:likelihood}
    \mathcal{L}(\mathrm{obs}_i | \alpha, \tmin, \tmax) \approx 
    \frac{\mathcal{A}}{N_\mathrm{samp}} \sum_{j=1}^{N_\mathrm{samp}}
    (\dot{n}_i^j \Delta t) e^{-\dot{n}_i^j \Delta t}
\end{equation}
where $\mathcal{A}$ is a constant that normalizes the likelihood, and $\Delta t$ is the fiducial observation time, which we assume to be $\Delta t = 10~\mathrm{yr}$, though this choice does not impact results so long as the expected number of events during the observing period satisfies $\dot{n} \Delta t \ll 1$ event, which is the case for all \ac{GRB} host galaxies in our sample. 
We note that this formulation of the hyperlikelihood assumes that the \ac{DTD} parameters are not prone to selection effects, that is, the probability of a detection does not depend on $\alpha$, \tmin, and \tmax. 
Selection effects may impact the detection of more highly offset (i.e., ``hostless'') systems with less luminous counterparts and/or weaker host associations, as well as systems at high redshifts with dimmer hosts; we discuss potential implications of selection effects further in Section~\ref{sec:discussion}. 

Assuming the observed short \ac{GRB} observations are independent, the hyperposterior for the \ac{DTD} parameters is 
\begin{multline}\label{eq:posterior}
    P(\alpha, \tmin, \tmax | \mathrm{obs}) \propto 
    \prod_{i=1}^{N_\mathrm{GRB}} 
    \mathcal{L}(\mathrm{obs}_i | \alpha, \tmin, \tmax) \\
    \times \pi(\alpha,\tmin,\tmax),
\end{multline}
where $N_\mathrm{GRB}$ is the number of \ac{GRB} hosts used in our sample, and $\pi(\alpha,\tmin,\tmax)$ is the prior distribution on the \ac{DTD} parameters. 
We assume that the priors on $\alpha$, $\tmin$, and $\tmax$ are independent and uniform on the ranges of $[0,3]$, $[5~\mathrm{Myr},500~\mathrm{Myr}]$, and $[3~\mathrm{Gyr},100~\mathrm{Gyr}]$, respectively. 

Sampling the likelihood of Equation~\ref{eq:likelihood} is computationally prohibitive, as it involves calculating an integral at each step in the parameter space of the \ac{DTD} distribution. 
To avoid this, we precompute a regular grid of likelihoods for each host galaxy and create an interpolant of these grids when evaluating the likelihood in Equation~\ref{eq:likelihood}. 
This grid contains $20$ points for each \ac{DTD} parameter uniformly spaced across their prior range, resulting in $8000$ likelihood interpolants for each host galaxy. 
We find that this approach provides consistent results compared to directly evaluating the likelihood, as the likelihood surface varies smoothly across the \ac{DTD} parameter space. 
For generating posterior distributions of the \ac{DTD} parameters, we use the \texttt{dynesty} nested sampler~\citep{Speagle2020} as implemented in \texttt{Bilby}~\citep{Ashton2019}.

\begin{figure*}[t]
\includegraphics[width=0.97\textwidth]{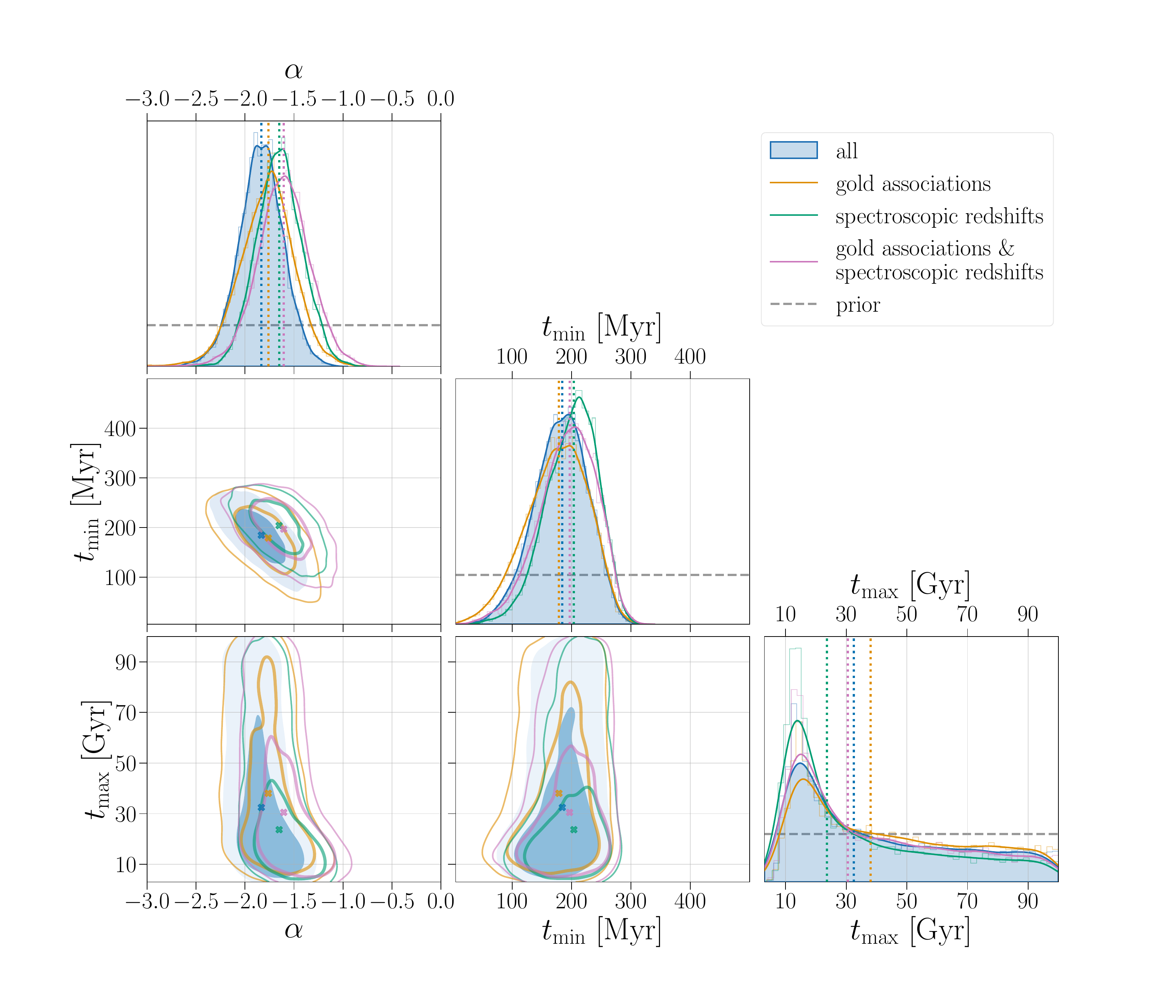}
\caption{Constraints on \ac{DTD} parameters when including all host associations with $\Pcc \leq 0.20$ in the inference (filled blue), as well as three subsets that only include samples that have $\Pcc \leq 0.02$ (yellow; Gold Association), samples that have spectroscopic redshifts (green), and samples that have spectroscopic redshifts and $\Pcc \leq 0.02$ (pink). 
The contours show the 50\% and 90\% credible regions, with dotted lines marking the median for each marginalized distribution. 
The gray dashed lines in the panels plotting the marginal distributions show the prior distribution for each parameter, which is uniform for all parameters of the \ac{DTD}. 
Both kernel density estimates and binned histograms are shown for the marginal distributions. 
}
\label{fig:cornerplot}
\end{figure*}

We note that this approach only considers host galaxies in which a short \ac{GRB} has been observed and not galaxies in which no short \ac{GRB} occurred during the fiducial observing period. 
Locally, this has been shown to lead to only a mild bias in the inferred \ac{DTD} parameters~\citep{Safarzadeh2019d}.

\subsection{DTD Constraints}\label{subsec:DTD_constraints}

Our main results are in Figure~\ref{fig:cornerplot}, which shows the constraints on the \ac{DTD} parameters from our hierarchical inference of Section~\ref{subsec:inference}. 
We show the posterior distributions of the \ac{DTD} parameters using the full population, as well as the three subsets of the population described in Section~\ref{subsec:criteria_host_associations}. 

Considering our full sample of \NHost \acp{GRB} in the inference, we find that the posterior distribution of \ac{DTD} parameters significantly deviates from the prior, with a power-law slope of ${\alpha = \AllSamplesAlpha}$ and a minimum delay time of ${\tmin = \AllSamplesTmin}$, where we quote the median and 90\% symmetric credible interval. 
These parameters, as expected, exhibit a strong correlation as the bulk of the short \ac{GRB} sample is consistent with relatively short delay times of $\mathcal{O}(100~\mathrm{Myr})$; higher values of $\alpha$ (i.e., a shallower power-law slope) correspondingly decrease the inferred values for $\tmin$ so that the \ac{DTD} probability distribution function still has enough support at the low end of the distribution to adequately explain the population as a whole. 

The constraints on the \ac{DTD} parameters are fairly consistent when using the full sample of \ac{GRB} hosts as opposed to the subsamples described in Section~\ref{subsec:criteria_host_associations}. 
When only using samples with $\Pcc \leq 0.02$ (i.e., Gold Associations) in the inference, the constraints on \ac{DTD} parameters show little variation compared to the full sample where hosts only satisfy $\Pcc \leq 0.20$. 
This suggests that the \ac{DTD} constraints are not strongly dependent on the specific criteria used for host associations, namely, the optical magnitude of the host and the offset of the \ac{GRB} with respect to the host. 
On the other hand, when considering only associations for which the host galaxy has a spectroscopically measured redshift, we find mild changes in our inferred \ac{DTD} parameters. 
In particular, the minimum delay time pushes to slightly larger values and becomes more tightly constrained with a median value that is $\DifferenceAllSamplesSpeczTminApproximate$ larger than when all hosts are included, though still within the range of uncertainty. 
Furthermore, the subsample with spectroscopic redshifts strongly excludes extremely short delay times, with $\tmin > \SpeczSamplesTminFirstPercentile$ at 99\% credibility. 
The power-law index shifts to slightly shallower slopes as opposed to the full sample, with $\alpha = \SpeczSamplesAlpha$. 
As the spectroscopic sample has precise redshift measurements, the modeled \acp{SFH} are much less uncertain compared to the photometric redshift sample, which is the main driver of these tighter constraints. 
We also note that most of the host galaxies in the spectroscopic redshift sample have redshifts $z<1$~\citep{Fong2022}, which may lead to mild differences in our \ac{DTD} constraints relative to the full sample as discussed in Section~\ref{sec:discussion}.

\begin{figure}[t]
\includegraphics[width=0.48\textwidth]{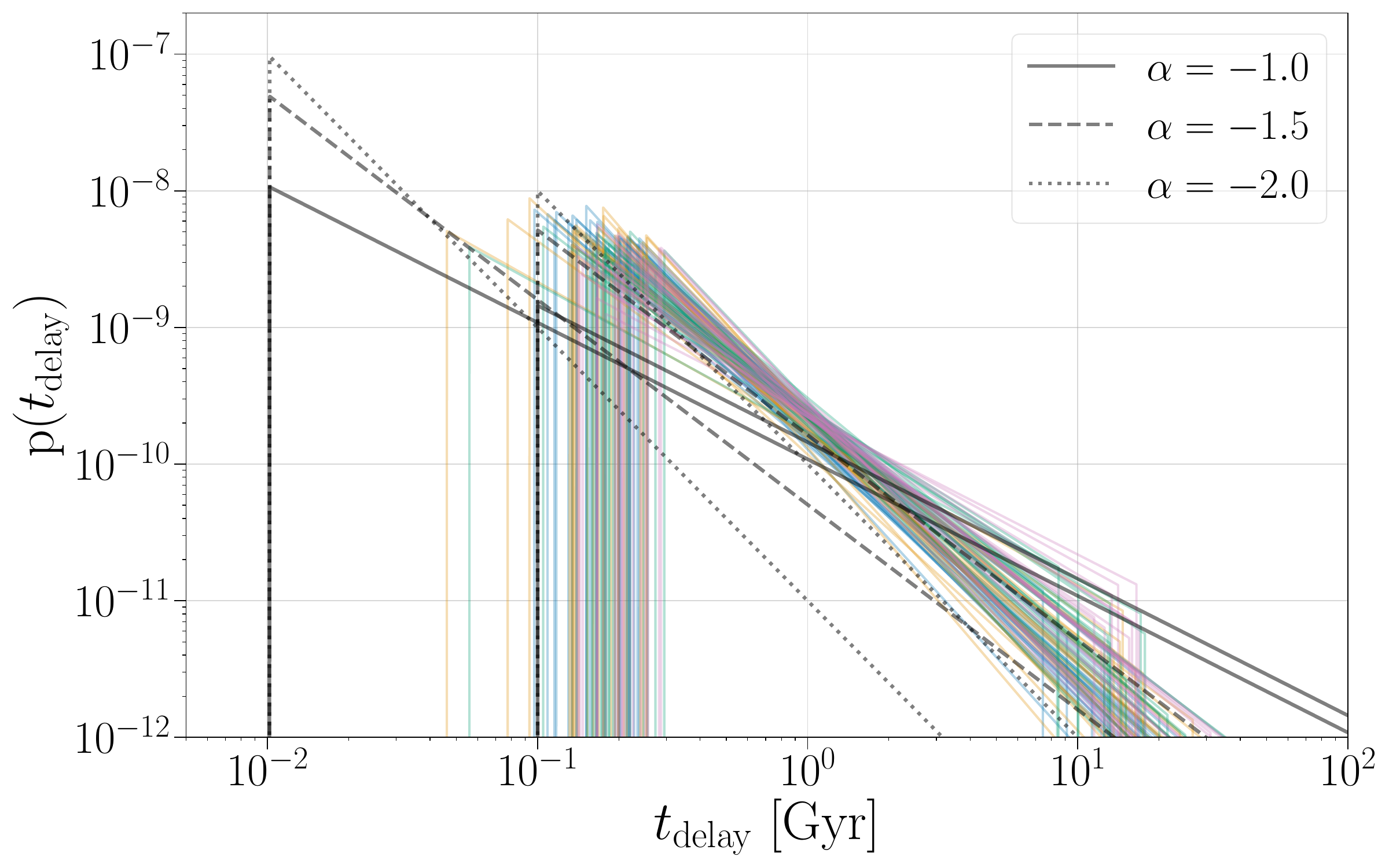}
\caption{\acp{DTD} constructed from random draws of our posterior distribution. 
The colored lines match the set of samples used in the inference as in Figure~\ref{fig:cornerplot}. 
The black solid, dashed, and dotted lines show fiducial \acp{DTD} with power-law slopes $\alpha$ of $-1$, $-1.5$, and $-2$, respectively, for comparison. 
We show this set of fiducial distributions for $\tmin = 10~\mathrm{Myr}$ and $\tmin = 100~\mathrm{Myr}$, with $\tmax$ fixed at $100~\mathrm{Gyr}$. 
}
\label{fig:DTD_draws}
\end{figure}

Neither $\alpha$ nor $\tmin$ are strongly correlated with $\tmax$, though for $\tmax \lesssim 30~\mathrm{Gyr}$, decreasing the value of $\tmax$ leads to slightly more support for smaller values of $\tmin$ and larger values of $\alpha$ (i.e., shallower power-law slopes). 
The posterior distribution for $\tmax$ peaks at \TmaxExtendedPeakApproximate, and the upper limit for $\tmax$ rails against the upper bound of our prior. 
Thus, we cannot make meaningful statements for the upper limit of $\tmax$, which translates to a lack of constraints on the maximum orbital separation of compact object binary systems containing a neutron star at formation. 
Posterior support for values below $\tmax \sim 10~\mathrm{Gyr}$ drops precipitously; using all host galaxies, we constrain $\tmax > \AllSamplesTmaxFirstPercentile$ at 99\% credibility. 
This constraint is driven by the systems in our population associated with quiescent host galaxies and old stellar populations. 

In Figure~\ref{fig:DTD_draws}, we show \acp{DTD} constructed from our posterior distributions, as well as \acp{DTD} using fiducial values for comparison. 
As described above, we find a preference for minimum delay times of $\tmin \sim 150~\mathrm{Myr}$ and power-law slopes of $\alpha \sim -1.5$. 
These constraints can help inform binary evolution modeling of \ac{BNS} progenitors, as slopes with $\alpha < -1$ may be indicative of hardening phases in the evolution of the progenitor such as mass transfer or common envelope phases~\citep{Belczynski2018}, and larger minimum delay times hint at proposed fast-merging \ac{BNS} channels (e.g., a Case BB common envelope scenario, in which the progenitor of the second-born neutron star proceeds through a second (unstable) mass transfer episode as an evolved naked helium star; see \citealt{Dewi2002,Ivanova2003}) may operate inefficiently.

\begin{figure}[t]
\includegraphics[width=0.48\textwidth]{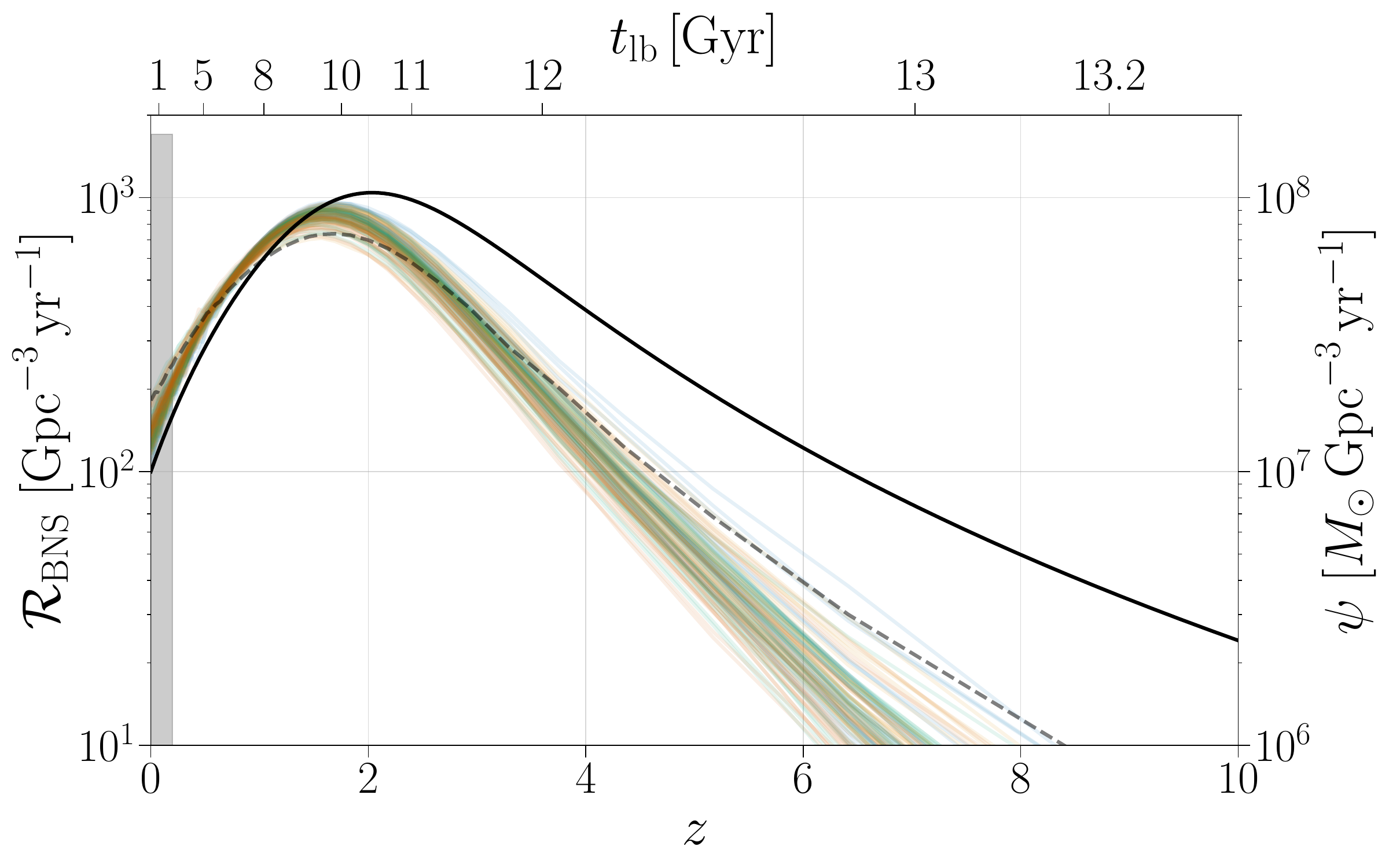}
\caption{Expected merger rate evolution of \acp{BNS} based on our \ac{DTD} constraints. 
The black line (right axis) shows the \ac{SFR} density evolution from \cite{Madau2017a}. 
The colored lines (left axis) match the set of samples used in the inference as in Figure~\ref{fig:cornerplot} and show the predicted \ac{BNS} comoving merger rate density for random draws from our \ac{DTD} parameter inference, assuming their progenitor stars are born according to the \ac{SFR} density of \cite{Madau2017a}. 
The black dashed line shows the expected \ac{BNS} merger rate evolution assuming fiducial \ac{DTD} values of $\alpha=-1$, $\tmin=10~\mathrm{Myr}$, and $\tmax=14~\mathrm{Gyr}$. 
The shaded gray region shows the local \ac{BNS} merger rate density constraints from \cite{GWTC3_pops}. 
}
\label{fig:merger_rate}
\end{figure}

\section{Implications for Gravitational Wave Observations}\label{sec:merger_rate_evolution}

Given our constraints on the parameters of the short \ac{GRB} \ac{DTD}, we can construct the expected merger rate evolution of \ac{BNS} mergers under the assumption that all short \acp{GRB} are the result of this class of compact object binary merger. 
We show the expected merger rate density evolution in Figure~\ref{fig:merger_rate}, assuming that \acp{BNS} follow the underlying cosmic \ac{SFH} of \cite{Madau2017a}. 
The \ac{BNS} merger rate density peaks at a redshift of $z \sim \MergerRatePeakRedshiftApproximate$. 
The overall normalization of the merger rate evolution is directly related to the assumed \ac{BNS} production efficiency, which we take to be fixed at $10^{-5}~\Msun^{-1}$. 
However, this assumption does not affect the shape of the merger rate evolution or the location of its peak. 
Given this fixed value for the \ac{BNS} production efficiency, the local merger rate density is predicted to be $\sim \LocalMergerRateApproximate$, consistent with the current constraints from the LIGO Scientific and Virgo Collaboration~\citep{GWTC3_pops}. 

The inferred \ac{BNS} merger rate evolution does not vary significantly for the different subsets of samples described in Section~\ref{subsec:criteria_host_associations} as the inferred \ac{DTD} is relatively robust. 
Though the redshift of the peak merger rate density is well beyond the \ac{BNS} horizon of current-generation \ac{GW} detectors operating at design sensitivity~\citep{LVC_ObservingScenarios}, it should be within the range of planned third-generation \ac{GW} detectors~\citep{Evans2021}. 
If inconsistencies are found between the peak of the \ac{BNS} merger rate density inferred from short \acp{GRB} and the peak of the \ac{BNS} merger rate observed by future \ac{GW} detectors, it may indicate an additional mechanism for producing short \acp{GRB} is at play or that other transients (e.g., long \acp{GRB}) are also caused by compact object binary mergers. 

\section{Discussion}\label{sec:discussion}

Although the \ac{DTD} constraints of short \acp{GRB} in this work are not dependent on the specific type of formation mechanism, these constraints can help inform uncertain aspects of massive-star binary evolution within the \ac{BNS} paradigm for short \acp{GRB}. 
Assuming that the merger time is purely driven by \ac{GW} inspiral (e.g., in the case of isolated binary evolution), the \ac{DTD} encodes information about the orbital properties of compact object binary systems at formation. 
Given the short stellar evolutionary timescales of \ac{BNS} progenitors prior to the second supernova that forms the compact object binary ($\sim 10$--$50~\mathrm{Myr}$; \citealt{Andrews2019}), the delay time is approximately the inspiral time, except for the potential low-end portion of the \ac{DTD} where a short inspiral time becomes comparable to the stellar evolution timescale. 

The \ac{DTD} can thus be used to place constraints on the minimum separation of \acp{BNS} at formation, as well as evolutionary phases in the progenitors that can steepen the compact object binary orbital separation distribution and therefore the \ac{DTD}~\citep{Belczynski2018,Broekgaarden2022a}. 
For example, given the first percentile of our recovered \tmin distribution when using all host galaxies ($t_\mathrm{min}^{1\%} = \AllSamplesTminFirstPercentile$) and assuming a \ac{BNS} formation timescale of $t_\star = 30~\mathrm{Myr}$, the minimum orbital separation at \ac{BNS} formation is $\approx \MinimumSeparationTminFirstPercentile$ (note that eccentricity in the orbit at \ac{BNS} formation would expedite the inspiral, leading to larger orbital separations for a given inspiral time). 
The preference of our results for longer minimum delay times $\gtrsim 100~\mathrm{Myr}$ may indicate that late-stage hardening phases, such as the Case BB common envelope scenario~\citep[e.g.,][]{Dewi2002,Ivanova2003}, may operate inefficiently or be nonexistent. 
This would have implications for the ability of \ac{BNS} systems to provide the $r$-process enrichment observed in low escape velocity environments such as ultra-faint dwarf galaxies~\citep{Safarzadeh2018} or environments with rapid star formation episodes such as globular clusters~\citep{Zevin2019b}, as enrichment in these environments requires short delay times of $\mathcal{O}(10~\mathrm{Myr})$. 
In addition to informing the distribution of orbital properties at compact object binary formation, delay times can be used to place joint constraints on the properties of a binary prior to the second supernova and the strength of the natal kick that formed the second neutron star~\citep[e.g.,][]{Andrews2015,Andrews2019a,Tauris2017}. 
Combining this information with the physical offset of the \ac{GRB} with respect to its host galaxy can further improve such constraints~\citep[e.g.,][]{GW170817_progenitor,Zevin2020c}. 
We will explore the constraints that this catalog places on \ac{BNS} progenitor pre-supernova orbital properties and natal kicks based on delay times and offsets in future work. 

A significant amount of work has explored constraints on \ac{BNS} progenitor properties based on the $\sim 20$ \ac{BNS} systems observed in the Milky Way~\citep{Andrews2015,Tauris2017,Andrews2019a,Andrews2019}. 
The Galactic \ac{BNS} population has uncertain selection effects that can be complemented with short \ac{GRB} host galaxy associations (see below for caveats on the short GRB sample selection). 
In particular, the shortest delay time \ac{BNS} systems in the Milky Way may be missed from surveys due to Doppler shifting from the orbit making their detection difficult, as well as their short lifetimes making their existence at any given point in time relatively rare~\citep{Tauris2017}. 
On the other hand, long-delay-time \ac{BNS} systems in the Milky Way may be hidden from surveys due to their long orbital periods, making their proper motion on the sky minuscule. 
We find a minimum delay time that is consistent with the shortest delay time systems in the Milky Way. 
For example, the \ac{BNS} system J1757-1854 is estimated to have one of the shortest delay times of the population observed in the Milky Way; it will merge in $\sim 70~\mathrm{Myr}$ and has a characteristic age of $\sim 130~\mathrm{Myr}$~\citep{Andrews2019a} based on the spin-down of the pulsar in the system, giving it a delay time of $\sim 210\mbox{--}250~\mathrm{Myr}$ with the inclusion of a $\sim 10\mbox{--}50~\mathrm{Myr}$ timescale between stellar birth and \ac{BNS} formation. 
This may be an indication that selection effects do not strongly impinge the detection of short-delay-time systems in the Milky Way, or that certain selection effects against observing short delay-time systems also affect the short \ac{GRB} population. 
The complementarity of these local and cosmological samples may also help us understand selection effects impacting the detection of counterparts to short \acp{GRB}, such as the possibility that highly offset short \acp{GRB} have fainter afterglows due to the tenuous nature of gas in the outskirts of galaxies~\citep[e.g.,][]{Berger2010,Tunnicliffe2014}. 

The largest uncertainty in our analysis likely lies in the reconstruction of the \ac{SFH}. 
Though the parametric delayed-$\tau$ \ac{SFH} has been shown to be an adequate representation of quiescent galaxies, other parametric or nonparametric approaches may be more physically realistic for \acp{SFH} of host galaxies that still have substantial star formation or bursty histories~\citep{Carnall2019,Leja2019}. 
However, due to the robustness of our results with differing subsets of \ac{GRB} hosts, we expect our main conclusions to hold despite this systematic uncertainty. 
In future work, we will explore how the choice of \ac{SFH} reconstruction affects our results by considering alternative parametric and nonparametric modeling techniques. 

The host galaxy of the multimessenger event GW170817/GRB170817, NGC4993, is quiescent with stellar age estimates of $\gtrsim3\mbox{--}10~\mathrm{Gyr}$~\citep{Blanchard2017,Im2017,Levan2017,Pan2017,Nugent2022}.  
Though more quiescent galaxies host short \acp{GRB} at low redshifts ($z < 0.5$) than at higher redshifts~\citep{Nugent2022}, properties of NGC4993 such as its mass-weighted age and specific \ac{SFR} make it a mild outlier relative to the other host galaxies used in this work (see \citealt{Nugent2022}). 
Despite this, the exclusion of GRB170817 in our analysis has a very minor impact on our \ac{DTD} constraints, only decreasing (increasing) our median recovered value of $\alpha$ (\tmin) by \SeventeenZeroEightSeventeenExclusionChange, well within the bounds of the uncertainty in these parameters. 
The \ac{DTD} constraints in this work are consistent with a long tail of delay times with $\tmax > \AllSamplesTmaxFirstPercentile$ at high credibility and thus can adequately account for the relatively old stellar population of NGC4993 and the long inferred delay time of GW170817~\citep{Blanchard2017,Adhikari2020}. 

Extremely short delay times of $\lesssim10~\mathrm{Myr}$ have been argued for the progenitor of GRB060505, which is spatially associated with an active star-forming region~\citep{Ofek2007}. 
This \ac{GRB} had a $T_{90}$ of $\sim4~\mathrm{s}$, longer than the typical delineation between short and long \acp{GRB} of $2~\mathrm{s}$. 
However, deep imaging ruled out the presence of a supernova (typically associated with long \acp{GRB}) to deep optical limits. 
Since $10~\mathrm{Myr}$ is approximately the shortest possible evolutionary timescale that a massive-star binary can form a \ac{BNS}, for a delay time of $\sim10~\mathrm{Myr}$ the inspiral timescale would need to be extremely short, requiring a \ac{BNS} birth semimajor axis of $a_\mathrm{BNS} \ll 1~\Rsun$ unless the \ac{BNS} was born with a high eccentricity. 
Such a scenario is inconsistent with our constraints on the minimum delay time $\tmin$; we find $\tmin > \AllSamplesTminFirstPercentile$ at 99\% credibility when using all host galaxies in our sample and $\tmin > \SpeczSamplesTminFirstPercentile$ at 99\% credibility when only considering host galaxies with spectroscopically measured redshifts. 
Given our \ac{DTD} constraints, this spatial coincidence with a star-forming region may instead be a lucky coincidence following the post-supernova migration of the \ac{BNS}, or this particular event may be a separate class of long \acp{GRB} without a supernova~\citep[e.g.,][]{Fryer2006}. 

However, a number of other long \ac{GRB} detections are lending credence to the possibility that some of these events are caused by compact object binary mergers. 
The discovery of a potential kilonova associated with the minute-long GRB211211A~\citep{Rastinejad2022} hints at \ac{BNS} mergers causing some fraction of the long \ac{GRB} population. 
Similarly, GRB060614 had a long duration (although is also classified as a possible short GRB with extended emission; \citealt{Lien2014}), and no coincident supernova was detected to deep optical limits~\citep{Gehrels2006}. 
As the properties of the host galaxies of GRB211211A and GRB060614 were also modeled in \cite{Nugent2022}, we performed additional analyses with the inclusion of these two hosts in our general short \ac{GRB} population. 
These provided nearly identical results. 
Even though our \ac{DTD} results are insensitive to the inclusion of these two long \acp{GRB}, the number of compact object binary mergers consistent with this picture of extended gamma-ray emission is uncertain and could potentially add more support for short delay times. 
However, it is unlikely that the majority of long \acp{GRB} result from a compact object merger paradigm, as many long \acp{GRB} in the local ($z < 0.5$) universe have been followed up extensively by electromagnetic observatories, and GRB211211A is the only one observed thus far to have a plausible kilonova counterpart~\citep{Rastinejad2022}. 

In addition to some fraction of \ac{BNS} mergers masquerading as long \acp{GRB}, our sample used to constrain the \ac{DTD} may suffer from other issues of incompleteness. 
As we rely on the modeling of host galaxies when constraining the \ac{DTD}, we do not consider short \acp{GRB} that do not have a confident host association. 
Though the properties of short \ac{GRB} hosts do not seem to deviate strongly as a function of the host-association confidence (see, e.g., Figure 4 of \citealt{Nugent2022}), neglecting these events may have a potential impact on both the low and high ends of our inferred \ac{DTD}. 
For example, \acp{GRB} that are highly offset from their hosts may have afterglows with much lower luminosities, making precise localization (and therefore host identification) difficult~\citep{Perna2022}. 
Such systems may have migrated over longer timescales to reach the highly offset locations of the burst and therefore may have longer delay times than the general population. 
Furthermore, the \Pcc method for host identification may incorrectly associate a \ac{GRB} with a faint underlying host rather than a bright host at a larger offset, though \citet{Fong2022} predicted this to be an effect only at the $\lesssim 7\%$ level. 
On the other hand, if such poorly associated \acp{GRB} are instead truly associated with faint galaxies that are below detection limits, we may be excluding additional systems with short delay times as these faint, low-mass galaxies are typically star-forming. 
Furthermore, though Swift can detect \acp{GRB} out to $z \sim 3$, there is likely some fraction of short \acp{GRB} that occur beyond this horizon, when the universe was $\lesssim2~\mathrm{Gyr}$ old. 
Short \acp{GRB} that occur at these early stages in the history of the universe must have short delay times, and this selection effect may bias the general population in our analysis to longer delay times.  
This would lead to a larger inferred \tmin, and, due to the correlation between \tmin and $\alpha$, more negative values of $\alpha$. 
However, this population of high-redshift short \acp{GRB} is likely small; assuming the \ac{SFH} from \cite{Madau2017a}, $<10\%$ of stars are born beyond $z=3$, and the fraction of compact object binary mergers beyond this redshift will be even smaller due to the delay time between formation and merger.

\section{Conclusions}\label{sec:conclusions}

We have placed constraints on the delay time distribution (\ac{DTD}) of short \acp{GRB} using the largest catalog of short \ac{GRB} host associations to date with the inclusion of inference on host galaxy properties. 
Assuming these transient events result from the merger of \ac{BNS} systems, the \ac{DTD} for short \acp{GRB} can be directly translated to the \ac{DTD} of \ac{BNS} mergers. 
This allows for predictions of the expected merger rate evolution, and can constrain aspects of massive-star binary evolution physics and compact object binary formation. 
Our main results are as follows: 
\begin{enumerate}
    \item Based on a catalog of \NHost short \ac{GRB} host galaxies, we constrain the \ac{DTD} of short \acp{GRB} to have a power-law slope steeper than flat-in-log, with a power-law index of $\alpha = \AllSamplesAlpha$, a minimum delay time of $\tmin = \AllSamplesTmin$, and a maximum delay time that is $>\AllSamplesTmaxFirstPercentile$. 
    \item Using different subsets of the full dataset that make cuts based on the host association confidence or whether the host galaxy has a spectroscopically measured redshift, we find our results to be robust. 
    However, when using the spectroscopic redshift sample (which is dominated by $z<1$ bursts), the \ac{DTD} pushes to slightly larger values of the minimum delay time and slightly shallower power-law slopes. 
    \item Assuming short \acp{GRB} are the result of \ac{BNS} mergers, we construct the expected merger rate evolution of \ac{BNS} mergers, which we predict to peak at a redshift of $z \sim \MergerRatePeakRedshiftApproximate$. 
\end{enumerate}

Constraints on the \ac{DTD} is the tip of the iceberg of what can be accomplished given a large sample of compact object binary host galaxies. 
This binary--host sample can enable novel constrains on properties of the compact object binary at birth, and pairing such inference with the physical offsets of short \acp{GRB} with respect to their hosts can help unveil supernova mechanisms and the strength of kicks that neutron stars receive at formation. 
Pairing this sample with multimessenger \ac{GW} events and the already existing population of \acp{BNS} in the Milky Way will help unravel uncertain selection effects and determine the ubiquity of \ac{BNS} mergers as the cause of short \acp{GRB}, all of which we aim to explore in future work. 
Even when multimessenger \ac{BNS} observations surpass the number of host-identified short \acp{GRB}, they will probe a lower redshift regime, and pairing these samples will be paramount for understanding the binary--host connection and its evolution over cosmic time.

\acknowledgments
We thank Jeff Andrews, Gourav Khullar, Joel Leja, and Vicky Kalogera for useful discussions. 
Support for this work and for M.Z. was provided by NASA through the NASA Hubble Fellowship grant HST-HF2-51474.001-A awarded by the Space Telescope Science Institute, which is operated by the Association of Universities for Research in Astronomy, Incorporated, under NASA contract NAS5-26555. 
A.E.N. acknowledges support from the Henry Luce Foundation through a Graduate Fellowship in Physics and Astronomy. 
W.F. gratefully acknowledges support by the David and Lucile Packard Foundation, the Alfred P. Sloan Foundation, and the Research Corporation for Science Advancement through Cottrell Scholar Award \#28284. 
D.E.H. is supported by NSF grants PHY-2006645 and PHY-2110507, as well as by the Kavli Institute for Cosmological Physics through an endowment from the Kavli Foundation and its founder Fred Kavli. 
This work was performed in part at the Aspen Center for Physics, which is supported by NSF grant PHY-1607611.

\software{\texttt{Astropy}~\citep{TheAstropyCollaboration2013,TheAstropyCollaboration2018}; 
\texttt{Bilby}~\citep{Ashton2019}; 
\texttt{iPython}~\citep{ipython}; 
\texttt{Matplotlib}~\citep{matplotlib}; 
\texttt{NumPy}~\citep{numpy,numpy2}; 
\texttt{Pandas}~\citep{pandas};
\texttt{Prospector}~\citep{Johnson2021};
\texttt{SciPy}~\citep{scipy}.}


\bibliography{library}{}
\bibliographystyle{aasjournal}

\end{document}